\documentclass[prl,aps,twocolumn,floatfix]{revtex4-1}
\usepackage{epsf}
\usepackage{epsfig}
\usepackage{graphics}
\begin{document}

\title{
Tunneling Spectroscopy Across the Superconductor-Insulator 
Thermal Transition} 
\author{Sabyasachi Tarat and Pinaki Majumdar}

\affiliation{Harish-Chandra  Research Institute,
 Chhatnag Road, Jhusi, Allahabad 211019, India}

\date{11 June 2014}

\begin{abstract}
Advances in scanning tunneling spectroscopy  reveal the presence of 
superconducting nanoregions well past the bulk thermal transition in 
strongly  disordered superconductors. We use a Monte Carlo tool to 
capture the spatially differentiated amplitude and phase fluctuations 
in such a material and establish spatial maps of the coherence peak 
as the superconductor is driven through the thermal transition. 
Analysis of the local density of states reveals that superconducting 
regions shrink and fragment with increasing temperature, but survive 
in small clusters to a temperature $T_{clust} \gg T_c$.  The gap (or 
pseudogap) in the spectrum survives in general to another independent 
scale, $T_g$, depending on the strength of interaction. This multiple 
scale description is consistent with recent measurements and defines 
the framework for analysing strongly disordered superconductors.
\end{abstract}

\maketitle

Although superconductors with $s$-wave symmetry are
robust to weak non-magnetic disorder
\cite{anderson,ag}, moderate
disorder can lead to large inhomogeneities in the
pairing amplitude, and strong  phase 
fluctuation between the `islands' that emerge.
This suppresses the transition
temperature, and beyond a critical disorder
the ground state becomes insulating
\cite{sit-revs}.
While the bulk features of the 
superconductor-insulator transition (SIT)
have been explored experimentally 
for several decades
\cite{haviland-prl,shahar-ovad,escoffier-2004,baturina-2007},
the recent use of high resolution scanning tunneling
spectroscopy (STS) 
\cite{sacepe-prl,sacepe-natc,sacepe-loc,pratap-prl,pratap-ph-diag,noat,pratap-th,sherman-2012}
has generated new questions 
about the superconducting state near the SIT.

The experiments allow two major advances. (i)~They
confirm the essentially inhomogeneous nature 
\cite{sacepe-prl,sacepe-natc,sacepe-loc,pratap-prl,pratap-ph-diag,noat,pratap-th,sherman-2012}
of the superconducting (SC) state, affirming that one does
 not have a homogeneous suppression of SC order with disorder
 and temperature.  
(ii)~They highlight the presence of {\it additional temperature scales}
in the problem, for example, a cluster formation scale,
$T_{clust}$, a pseudogap formation scale, $T_{pg}$, and, at strong disorder,
 a possible gap formation scale $T_g$ - all distinct
from $T_c$. 
In addition, STS measurements  quantify
the detailed behaviour of the local density of states (LDOS)
with disorder and increasing temperature
\cite{sacepe-loc,pratap-th,sacepe-prl,pratap-ph-diag,noat,sherman-2012}
  - posing a challenge for theories
 that address only average properties. 

Addressing these issues calls for an approach
 that captures the increasing fragmentation in the ground
 state and retains the crucial phase and amplitude fluctuations
 that dictate thermal properties. 
Mean field Hartree-Fock-Bogoliubov-de-Gennes (HFBdG) theory
\cite{ghosal-prb} reasonably describes the ground state
but ignores phase fluctuations, while quantum Monte Carlo 
(QMC) \cite{triv-1996,bouadim}
retains all fluctuations but lacks 
spatial resolution.
Using an auxiliary field scheme that 
captures the HFBdG ground state, and the
correct $T_c$ and critical disorder $(V_c)$, 
we provide a spatially resolved description of the
thermal transition and estimate the emergent scales 
in a strongly disordered superconductor.  

Working at moderate interaction ($U=2t$, see
later) we confirm the fragmentation of the
superconducting ground state with increasing 
disorder, with SC islands surviving in an 
`insulating' background.
Our key results on thermal behaviour of the LDOS 
are the following:
(i)~At weak disorder increasing temperature ($T$) 
leads to spatially 
homogeneous closure of the gap at $T_c$.
For $V \rightarrow V_c$ the $T=0$ gaps are lower
in the SC regions than in the insulator,
increasing $T$ reduces all gaps but they survive to
a scale $T_g \gg T_c$, and  
a pseudogap is observed to $T_{pg} \gg T_g$. 
(ii)~In the weakly disordered system
the coherence peak in the LDOS vanishes
throughout the system at $T=T_c$. At strong disorder 
it survives on isolated clusters to a scale $T_{clust} 
\gg T_c$.
(iii)~The scales $T_g$, $T_{clust}$, {\it etc}, 
have distinct physical origin. We establish 
their variation with disorder and interaction strength.   
Finally, (iv)~we suggest a simple lattice Ginzburg-Landau
model, with parameters extracted from the electronic
problem, that reasonably describes the complex
thermal behaviour.
%
\vspace{-.2cm}
\begin{figure}[b]
\centerline{
\includegraphics[width=7.5cm,height=7.5cm,angle=0]{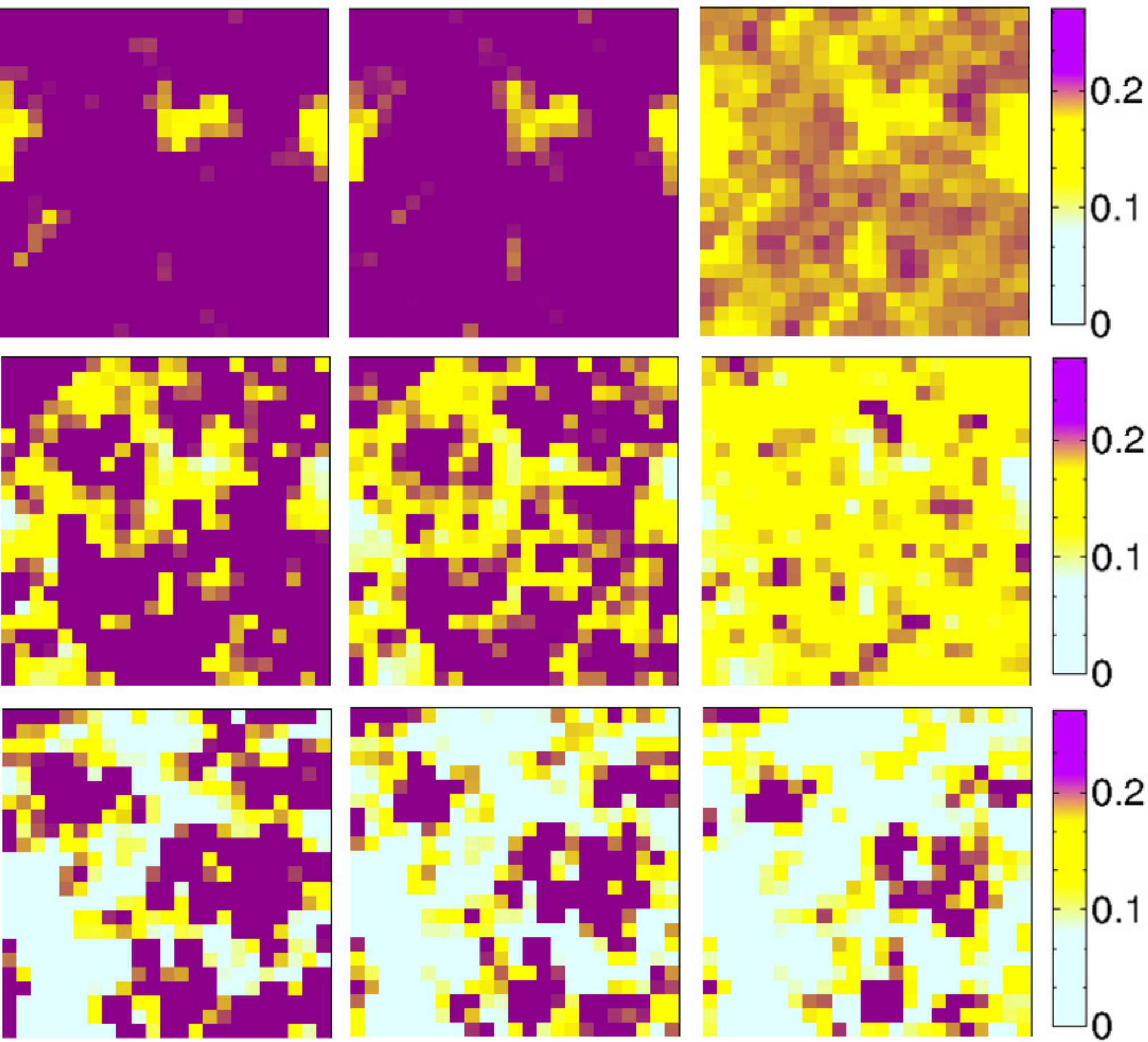}
}
\caption{Colour online: Maps of the tunneling conductance
integrated over a narrow frequency window around the coherence peak
feature in the LDOS (see text).
Rows, top to bottom, $V=0.2V_c, 0.5V_c, 0.9V_c$.
Columns, left to right, $T/T_c(V) = 0, 0.5, 1.0$.
Thermal average over 100 configurations.
}
\end{figure}

We study the attractive two dimensional 
Hubbard model (A2DHM) in the presence 
of potential scattering:
$
H = H_{kin}
+ \sum_{i \sigma} (V_i - \mu) n_{i \sigma}
- \vert U \vert \sum_{i} n_{i \uparrow} n_{i \downarrow}
$ with
$H_{kin} =
- t\sum_{\langle ij \rangle \sigma}
c_{i \sigma}^{\dagger} c_{j \sigma}$. 
$t$ is the nearest neighbour tunneling amplitude.
$V_i$ is a random potential picked from a normalised
flat distribution between $\pm V$. $\mu$ is the
chemical potential controlling the electron density $n$.
We fix $\mu$ so that $n \approx 0.9$.
$U > 0$ is the strength of onsite attraction.
We will set $t=1$ and measure all energies, and
temperature $(T)$, in units of $t$.
We set $U=2t$, to be close to
the experimentally relevant 
weak coupling window.

The difficulty of the A2DHM lies in handling the
interaction term.  We use a decomposition 
\cite{dubi,sit-transp,bcs-bec}
of the
interaction in terms of a pairing field, $\Delta_i =
\vert \Delta_i \vert e^{i \theta_i}$, and a density
field $\phi_i$ and treat these fields as classical.
This leads to the effective Hamiltonian 
$
H_{eff}  = H_{kin}  + \sum_{i \sigma} (V_i - \mu) n_{i \sigma}
+ H_{coup} + H_{cl}$,
where $H_{coup} = \sum_i (\Delta_i c^{\dagger}_{i\uparrow}
 c^{\dagger}_{i \downarrow}
+ h.c) - \sum_i \phi_i n_i$ and $H_{cl} =
{1 \over U} \sum_i (\vert \Delta_i \vert^2 + \phi_i^2)$.
We solve the coupled fermion-auxiliary
field problem through a Monte Carlo
\cite{dubi,dag,tca}. 
At finite $T$ this allows us 
to consider electron propagation in an amplitude and
phase fluctuating background, 
affording a dramatic improvement in the handling
of thermal physics.
We have discussed the
method in detail elsewhere \cite{sit-transp,bcs-bec} 
so we move on to the results.

The `clean $T_c$' at $U=2t$ is $T_c^0 \approx 0.07t$. 
Increasing
disorder pushes our $T_c$ below measurement resolution
($\sim 0.005t$) at $V \approx 2t$. We set this as $V_c$
\cite{vc-footnote}.
Based on the bulk transport and spectral properties, 
we characterise \cite{sit-transp} 
$V \lesssim 0.25V_{c}$ as `weak' disorder, 
$0.25V_{c} \lesssim V \lesssim 0.75V_{c}$ as intermediate,
and
$V \gtrsim 0.75V_{c}$ as strong disorder. 
The weak disorder regime is characterised
by a featureless DOS and metallic transport for $T > T_c$,
intermediate disorder involves a pseudogap (PG) 
for $T > T_c$ and a thermal crossover from 
insulating to metallic resistivity, while 
strong disorder involves a hard gap over a window
$T_g > T > T_c$ and activated transport at high $T$.

Fig.1 presents a summary of the thermal evolution of 
the coherence peak map 
at weak, moderate, and strong  disorder.
Our data shows the integrated tunneling conductance (TC) 
over the window $[\omega_c^-,\omega_c^+]$,
defined by $T_i^{coh}
= \langle \int_{\omega_c^-}^{\omega_c^+}
d\omega N_{ii}(\omega) \rangle$, where
$N_{ii}(\omega)$ is the local density of states at
site ${\bf R}_i$. $\omega_c^-=0.2t$ and $\omega_c^+ = 0.45t$
are chosen so that they cover
the 
coherence peaks in the global density of state, and hence
gives information about the local phase correlations in the system.

We make the following observations:
(a)~At weak disorder the pattern remains almost homogeneous
at all $T$, except for a few isolated
regions. Coherence peaks get suppressed with increasing 
temperature, and vanish by $T=T_{c}$.
(b)~At intermediate disorder 
the ground state is noticeably inhomogeneous 
and increasing $T$ causes further fragmentation.
However, by the time $T=T_c$ hardly any coherence peaks
are visible anywhere. 
(c) The high disorder regime shows tenuously connected clusters at $T=0$, 
which shrink as $T$ is increased, but have a prominently visible
but disconnected pattern at $T=T_c$. In fact at $V=0.9V_c$ the
clusters are visible to $T \sim 2T_c$. The `cluster survival scale'
at $V_c$ is 
$\sim 0.6T_c^0$ and drops slowly with increasing disorder.

\begin{figure}[b]
\centerline{
~~
\includegraphics[width=8.0cm,height=11.0cm,angle=0]{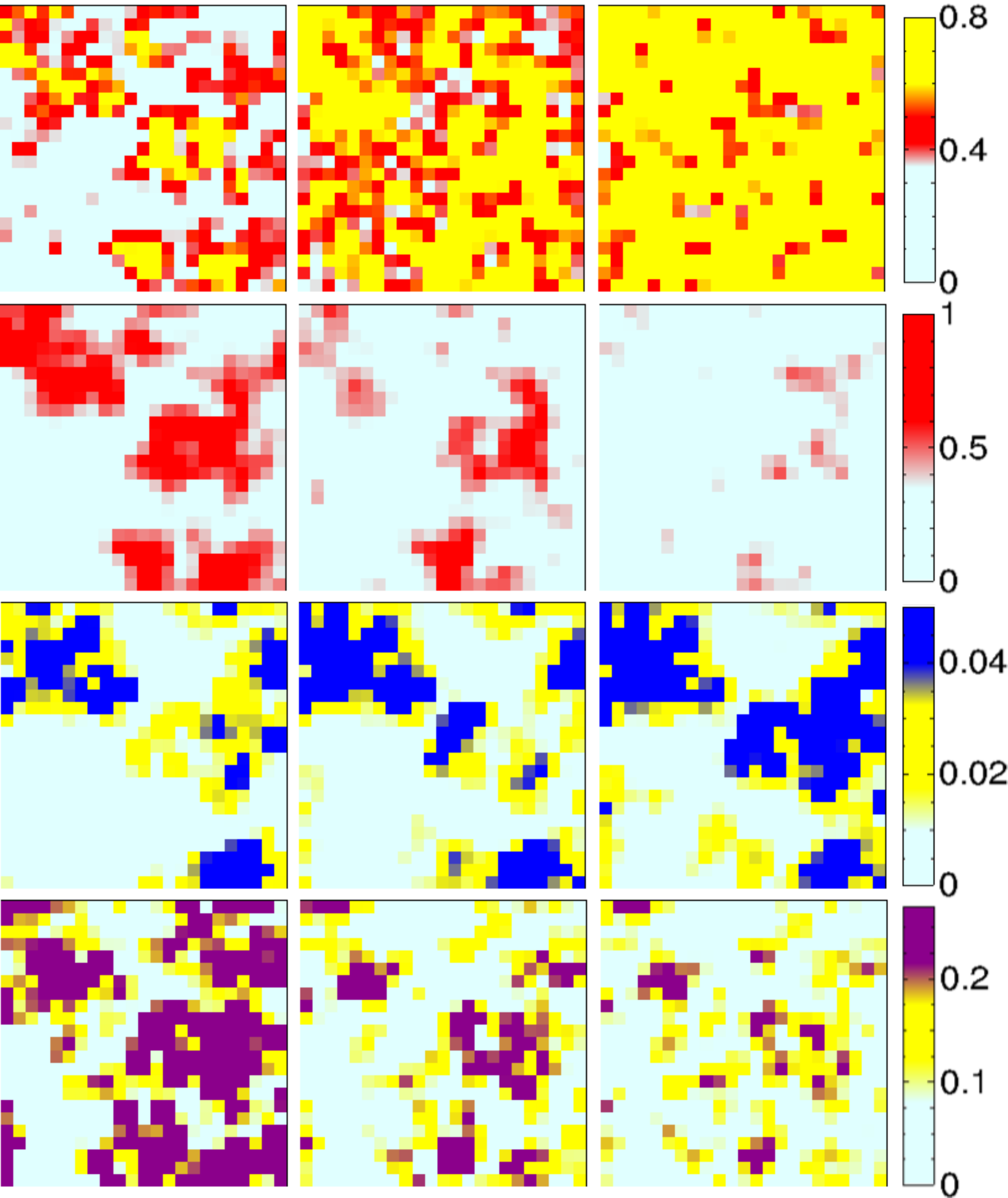}
}
\caption{Colour online: Spatial maps at $V=0.9V_c$.
1st row: 
$\langle \vert \Delta_i \vert \rangle$,
2nd row:
phase correlation $\Phi_i$,
3rd row:
tunneling conductance $T_i^{gap}$,
4th row: $T_i^{coh}$. 
The notation is explained in the text.
Columns, from left to right, are for
$T=0,~T_c,~2T_c$.
The interpretation of these patterns is discussed in the text.
}
\end{figure}

Fig.2 shows spatial maps of the pairing field and the tunneling
conductance,  averaged over 100 thermal configurations, at strong
disorder ($V= 0.9 V_c$) for a single realisation of disorder.
The Supplement shows results at weaker
disorder.
The top row shows 
$\langle \vert \Delta_i \vert \rangle$, normalised by 
the clean $T=0$ value $\Delta_0$.  Next row:
nearest neighbour averaged 
phase correlation $\Phi_i = 
\langle
{1 \over 4} \sum_{\delta}
cos(\theta_i -\theta_{i + \delta})
\rangle$, 
where $\delta$ refer to the four nearest neighbours of
a site. Third row: $T_i^{gap}=
 \langle 
\int_{\omega_g^-}^{\omega_g^+} d\omega
N_{ii}(\omega)
\rangle $,
the local 
tunneling conductance probed at subgap frequencies.
Fourth row: $T_i^{coh}$.
We set $\omega_g^-=0,~\omega_g^+ = 0.2t$.
Columns, left to right, correspond to
$T=0,~T_c,~2T_c$.

Let us start with the patterns at
$T=0$, left column.
(a).~We see
a clear separation between
regions where 
$\langle \vert \Delta_i \vert \rangle 
\gtrsim 0.4\Delta_{0}$, and where 
$\langle \vert \Delta_i \vert \rangle
\ll 0.4\Delta_0$.
While there is significant variation in magnitude {\it within}
the larger $\Delta$ regions, the distinction between
large and small $\Delta$ regions is unambiguous. 
(b).~The large $\Delta$ regions are phase correlated:
$\Phi_i$ is large in regions where 
$\langle \vert \Delta_i \vert \rangle
$ is large. These regions 
are the SC clusters.
(c).~$T_i^{gap}$
shows that the large
$\langle \vert \Delta_i \vert \rangle$
phase correlated regions have {\it large}
subgap TC, while regions with poor SC
correlation have virtually zero TC.
This suggests a smaller local gap in the
SC clusters, as we will confirm later,
and a larger gap in the non SC
regions.
The behaviour is in contrast 
to homogeneous systems where  larger 
$\langle \vert \Delta_i \vert \rangle$
would have meant a larger gap and a
{\it smaller} TC.
(d).~The map for  $T_i^{coh}$ shows
that the SC clusters in the ground state have
a modest coherence peak, while there is no CP
in the larger gap non SC regions.
Overall, at $T=0$ the non SC regions have no
noticeable spectral weight from
$\omega=0$ to frequencies
well beyond the average CP location.

Now the thermal evolution.
By $T=T_c$, middle column,
we observe the following.
(a)~There is significant homogenisation of  
$\langle \vert \Delta_i \vert \rangle$.
Non SC regions generate a 
strikingly large 
$\langle \vert \Delta_i \vert \rangle$ while the
SC clusters see a more modest growth from the
$T=0$ value. Temperature
leads to 
strong spatially differentiated
 amplitude fluctuation in the system. 
(b)~$\Phi_i$
shows thermal shrinking of the
correlated regions. It is 
still large in parts of the 
regions which had the large 
$\langle \vert \Delta_i \vert \rangle$ at $T=0$.
The clusters are
internally correlated but disconnected. The independent
fluctuation of the phase of the different clusters 
leads to loss of global SC order.
(c)~There is no noticeable change in 
$T_i^{gap}$ with $T$ for regions that were 
non SC at $T=0$. For SC regions there is an increase
in intensity. (d)~For $T_i^{coh}$, as we have
already seen in Fig.1, areas  with strong CP
feature shrink but are still clearly visible. Non SC
regions do not respond to temperature.

By $T = 2T_{c}$, 3rd column,
(a)~the mean magnitude 
has homogenised, with traces of clustering apparently lost,
and (b)~$\Phi_i$ is virtually zero everywhere.
The homogenisation of amplitude and phase variables may
suggest that any imprint of the $T=0$ 
cluster pattern would
be lost. However, (c)~the subgap TC is still
very inhomogeneous, but now uniformly
large over regions that were SC at $T=0$.
So, even at this ``high temperature'' the
subgap TC reveals the granularity of the ground state. 
Finally, (d)~the high intensity regions in
$T_i^{coh}$ shrink 
and the pattern tends towards a homogeneous
intermediate intensity with only small remnants of the
high CP regions. 

We have observed that at weaker disorder 
the correspondence of $T^{i}_{gap}$
with the background superconducting pattern weakens.
The data at $V=0.5V_{c}$ in the Supplement show 
that there is no clear correspondence between 
$T^{i}_{gap}$ and the 
correlated regions.
$T^{i}_{coh}$, however, continues to roughly
track the superconducting order, both in the ground state
and at finite $T$, down to low
disorder.

Fig.3 quantifies the distributions, $P(g)$ and $P(h)$,
 of gap magnitude and coherence peak
integral, respectively,
across the system.
The results are for $V= 0.2V_c$ and $0.9V_c$, and $T/T_c(V) = 
0,~0.5,~1.0$.
At $0.2V_c$ 
the $P(g)$ has a mean  $\approx  2 \Delta_{0}$ at $T=0$, 
with a narrow width around it. With
increasing $T$ the mean `gap' shifts to 
lower values while the
width shows a small increase. 
This suggests a 
homogeneous decrease throughout the system. 
At $0.9V_c$ $P(g)$ is wide at $T=0$ 
with a large gap tail arising from
sites with large positive or negative 
effective potential (we call these
hill and valley sites). Increasing $T$ leads to
shift in weight to lower $g$ from intermediate values
while $P(g)$ at large $g$ remains
unaffected.

\begin{figure}[t]
\centerline{
\includegraphics[width=4cm,height=3.5cm,angle=0]{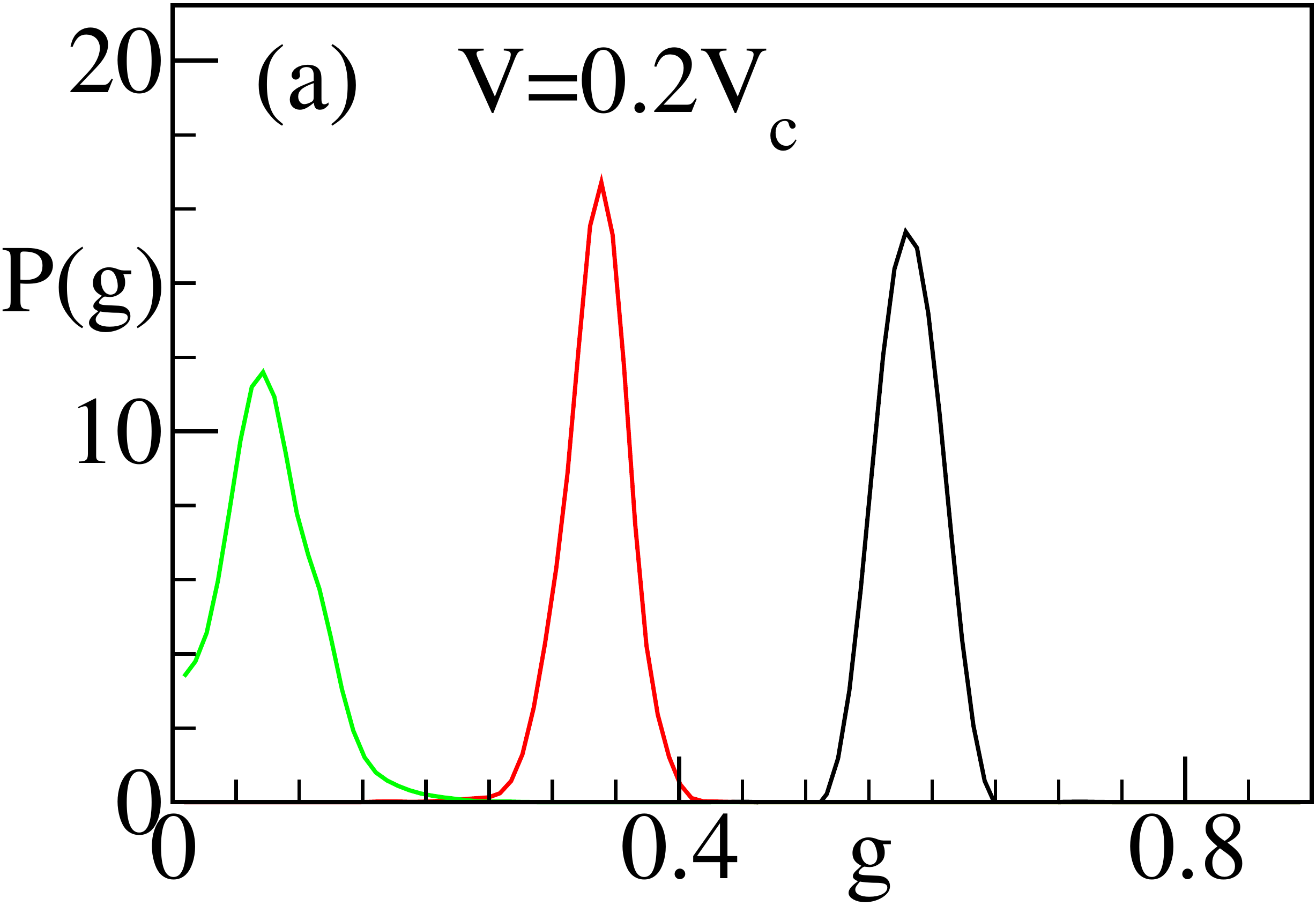}
\includegraphics[width=4cm,height=3.5cm,angle=0]{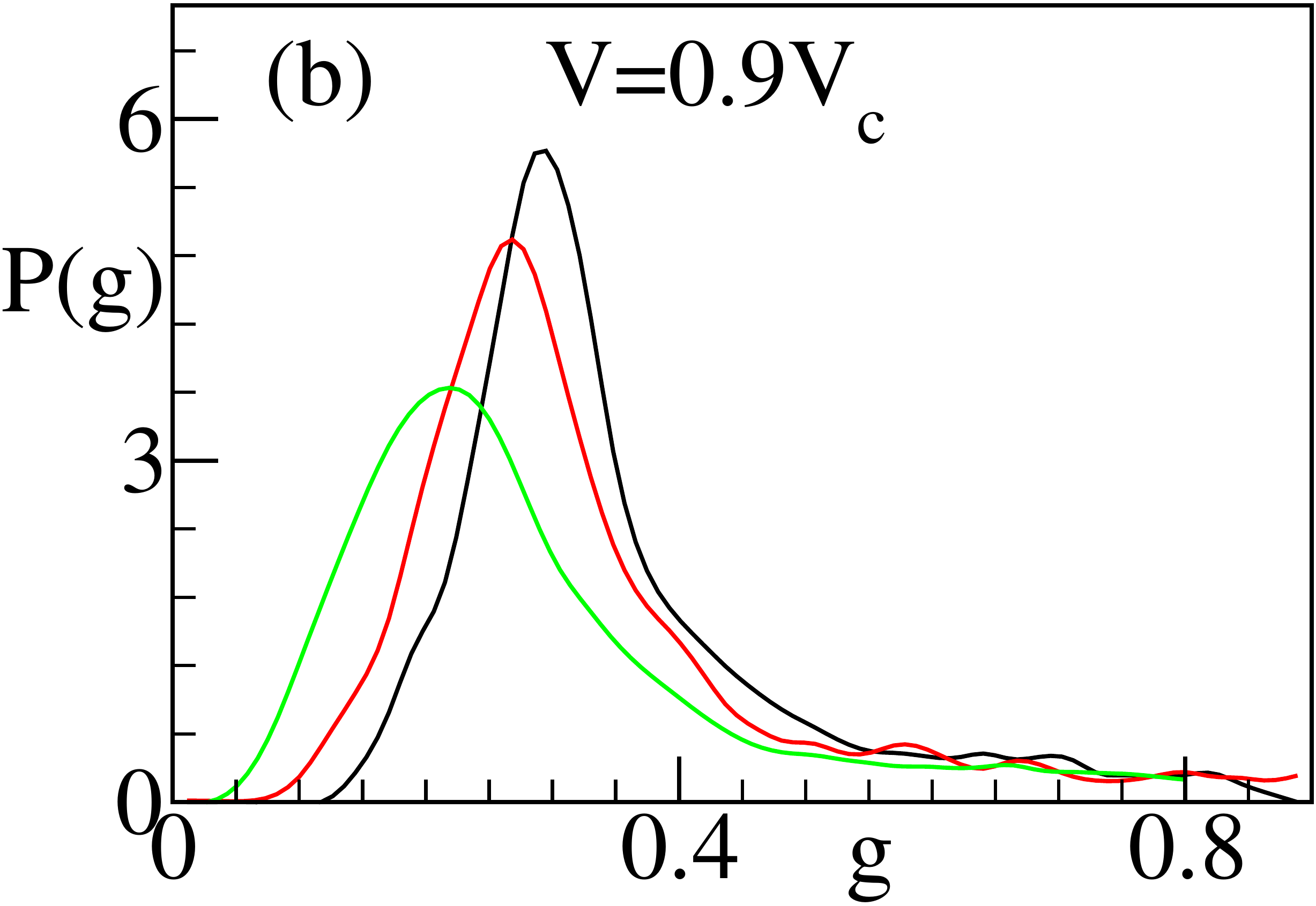}
}
\centerline{
\includegraphics[width=4.2cm,height=3.5cm,angle=0]{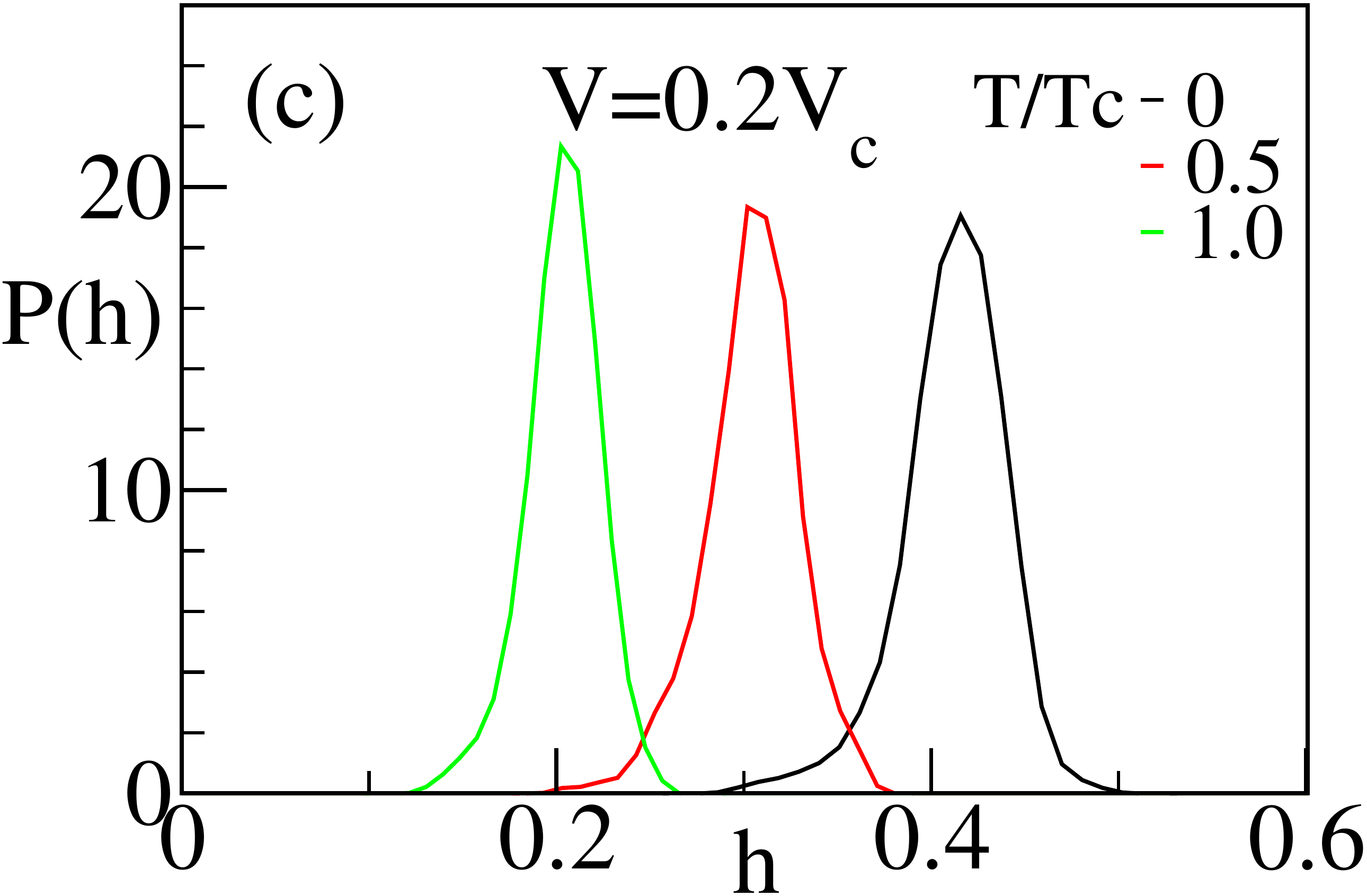}
\includegraphics[width=4.2cm,height=3.5cm,angle=0]{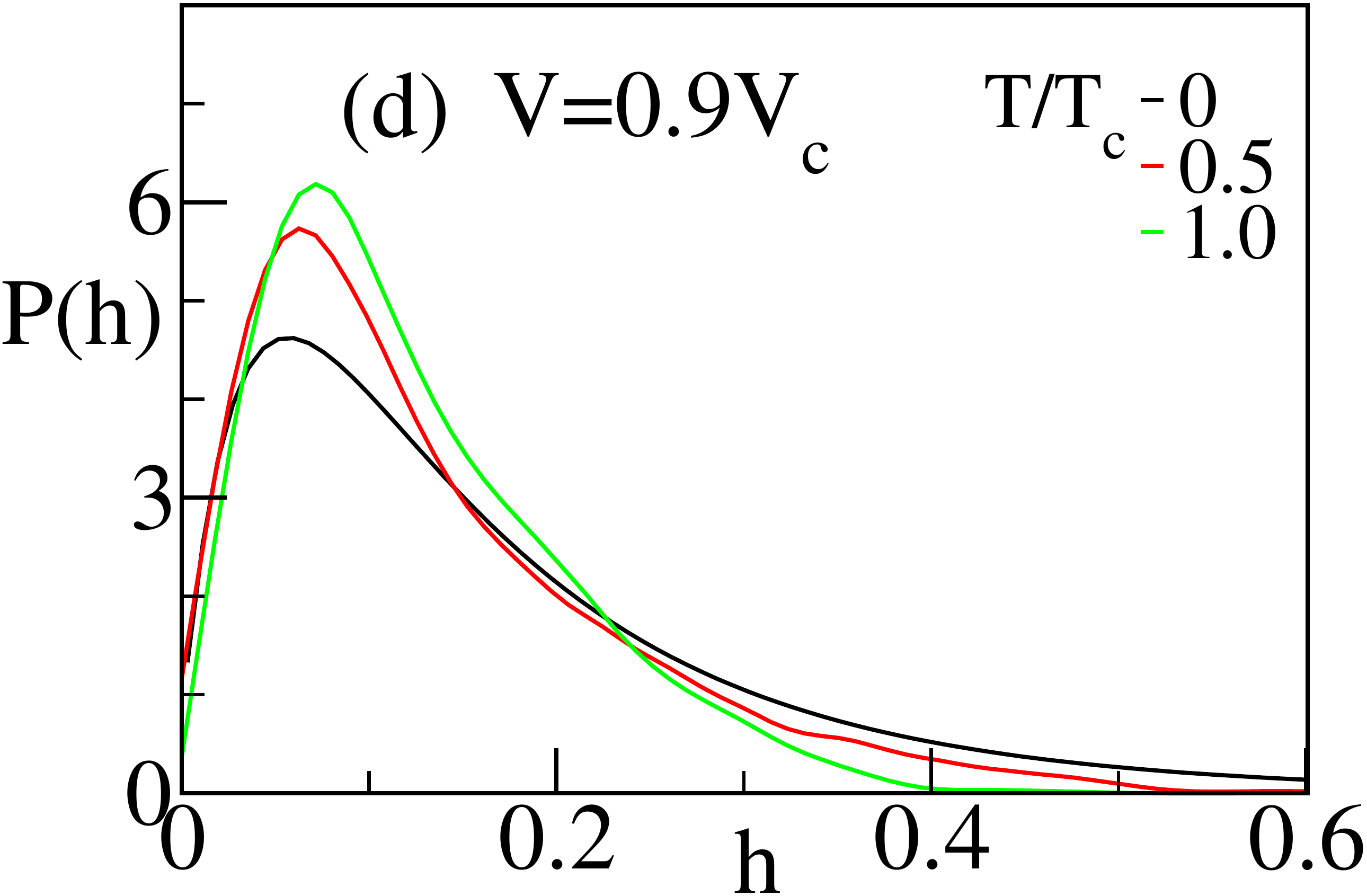}
}
\caption{Colour online:
Thermal evolution of the gap and coherence peak height distribution.
Top: Gap distribution for $V=0.2V_{c}$ (left) and $0.9V_{c}$ (right). 
Bottom: Coherence peak height  distribution at same $V$.
$T/T_c(V) = 0, 0.5, 1.0$.
}
\end{figure}

Coming to $P(h)$, panel (c) shows that 
the coherence peak distribution is also roughly
uniform at $0.2V_c$ at all $T$. 
The peak at $h \gtrsim 0.4$ at $T=0$
narrows slightly and moves to lower 
values at higher $T$ 
but the mean remains finite  
since we have not subtracted 
the high $T$  background.
At $0.9V_c$, however, most sites have poor 
coherence features, defining the trunk of the distribution,
except for the tail with $h \gtrsim 0.4$ 
arising from sites in the superconducting clusters.
With increasing $T$ as the SC regions shrink 
the weight in this $h \gtrsim 0.4$ region is lost.


Fig.4 correlates the low energy features of local gap and
coherence peak to spectral weight distribution over a
wider frequency window. We plot the LDOS at two
representative sites  (`plateau' 
and `hill') at low and high disorder. 
The plateau site involves an effective potential
$V_i - \phi_i$ close to the mean
value, and a local density $n_i$ close to the average,
$n_{av} \sim 0.9$,  while the hill
site has a large positive 
effective potential and $n_i \ll n_{av}$. 
At low disorder, Fig.4(a)-(b), both sites 
show coherence peaks and similar gaps, with the hill site
naturally having 
larger weight at $\omega > 0$.
The thermal evolution is also similar, with both gaps decreasing 
and closing at $T \lesssim T_c$.

At high disorder, Fig.4(c)-(d), the  LDOS at
the plateau site (part of a SC cluster) 
shows a narrow gap at low $T$, moderate coherence peaks, 
and expected thermal behaviour. The hill site, by contrast,
shows a large gap (strongly suppressed low frequency spectral
weight), no coherence peaks, and
is virtually insensitive to $T$.

\begin{figure}[t]
\centerline{
\includegraphics[width=4cm,height=3.4cm,angle=0]{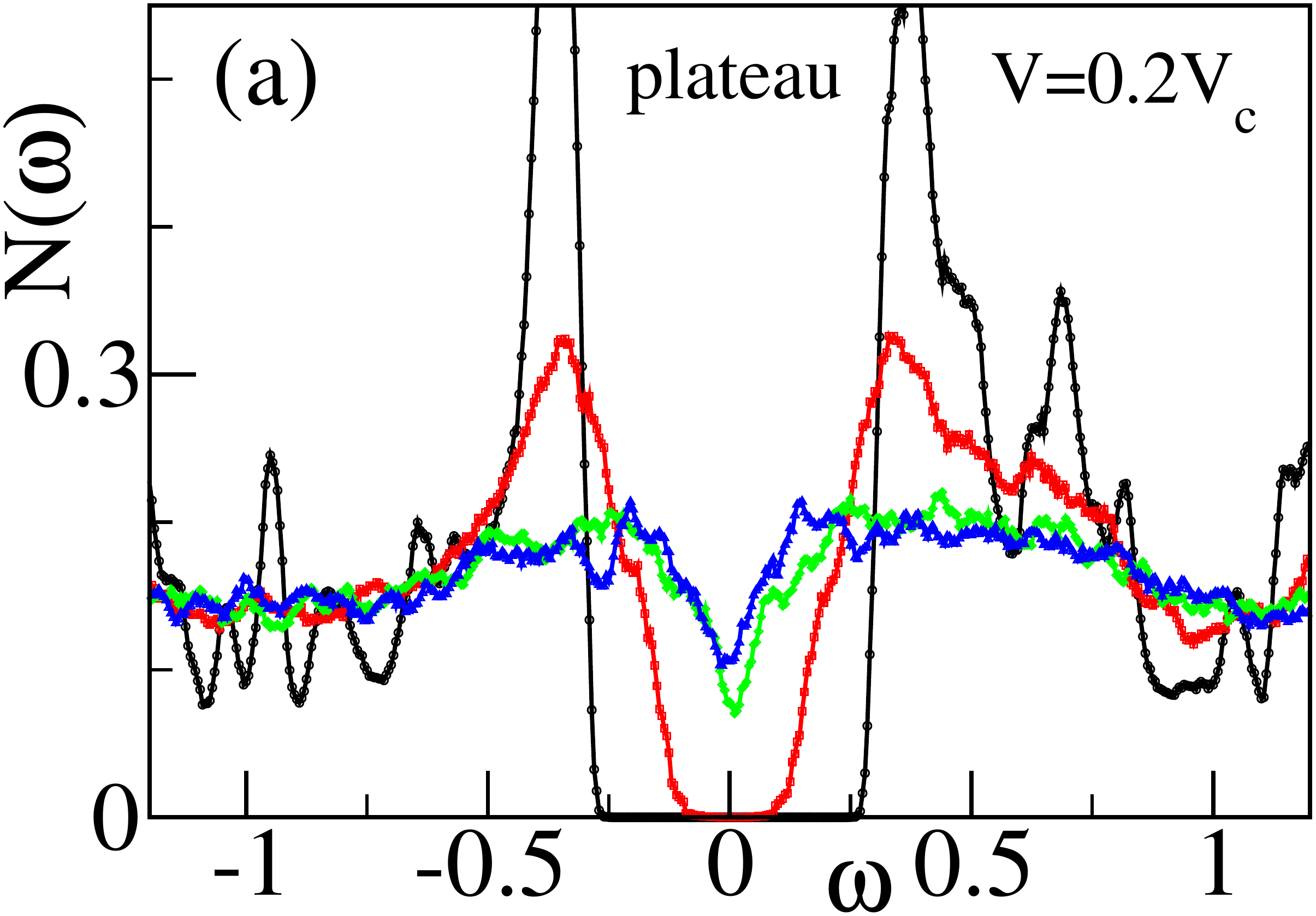}
\includegraphics[width=4cm,height=3.4cm,angle=0]{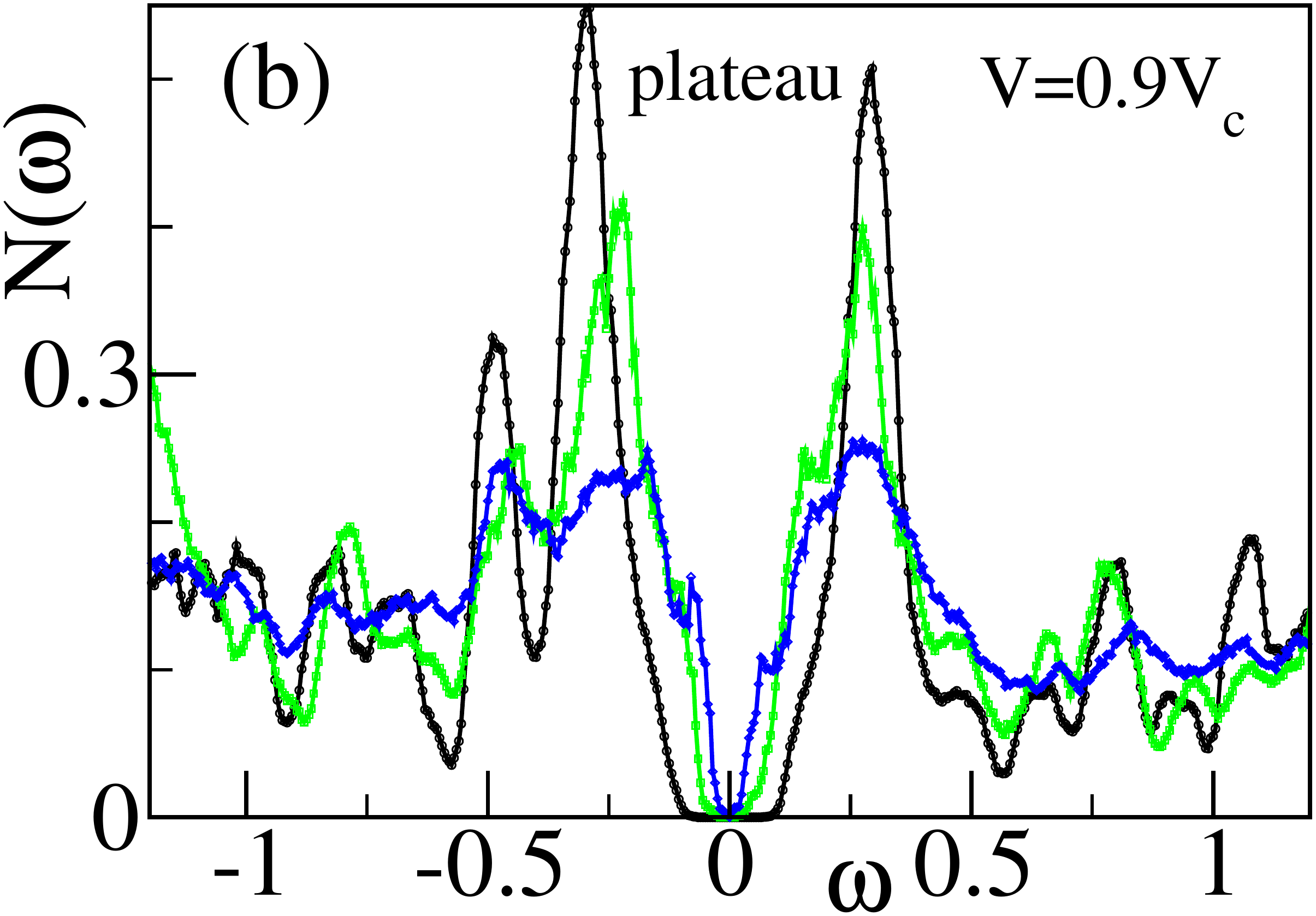}
}
\vspace{.2cm}
\centerline{
\includegraphics[width=4cm,height=3.4cm,angle=0]{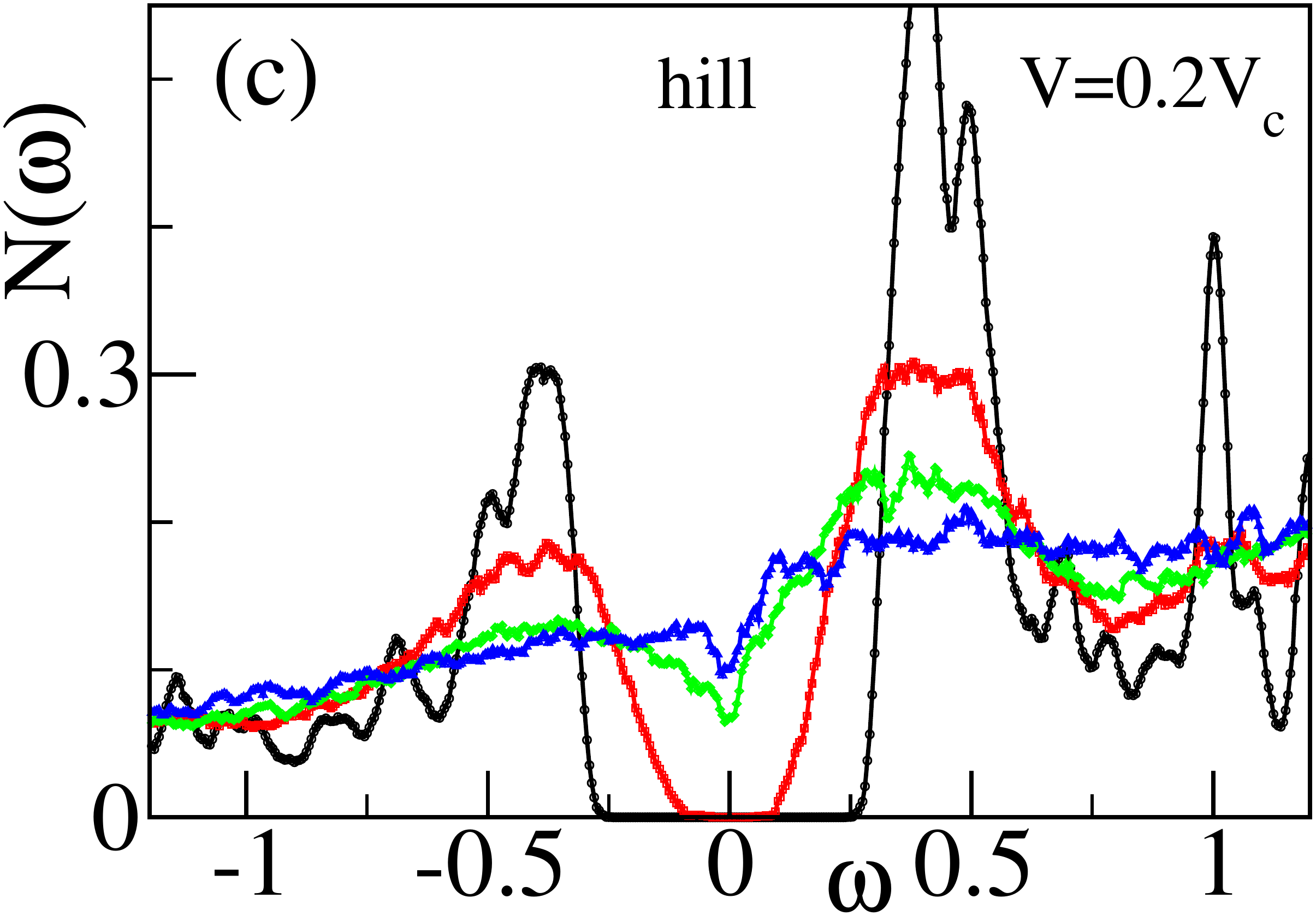}
\includegraphics[width=4cm,height=3.4cm,angle=0]{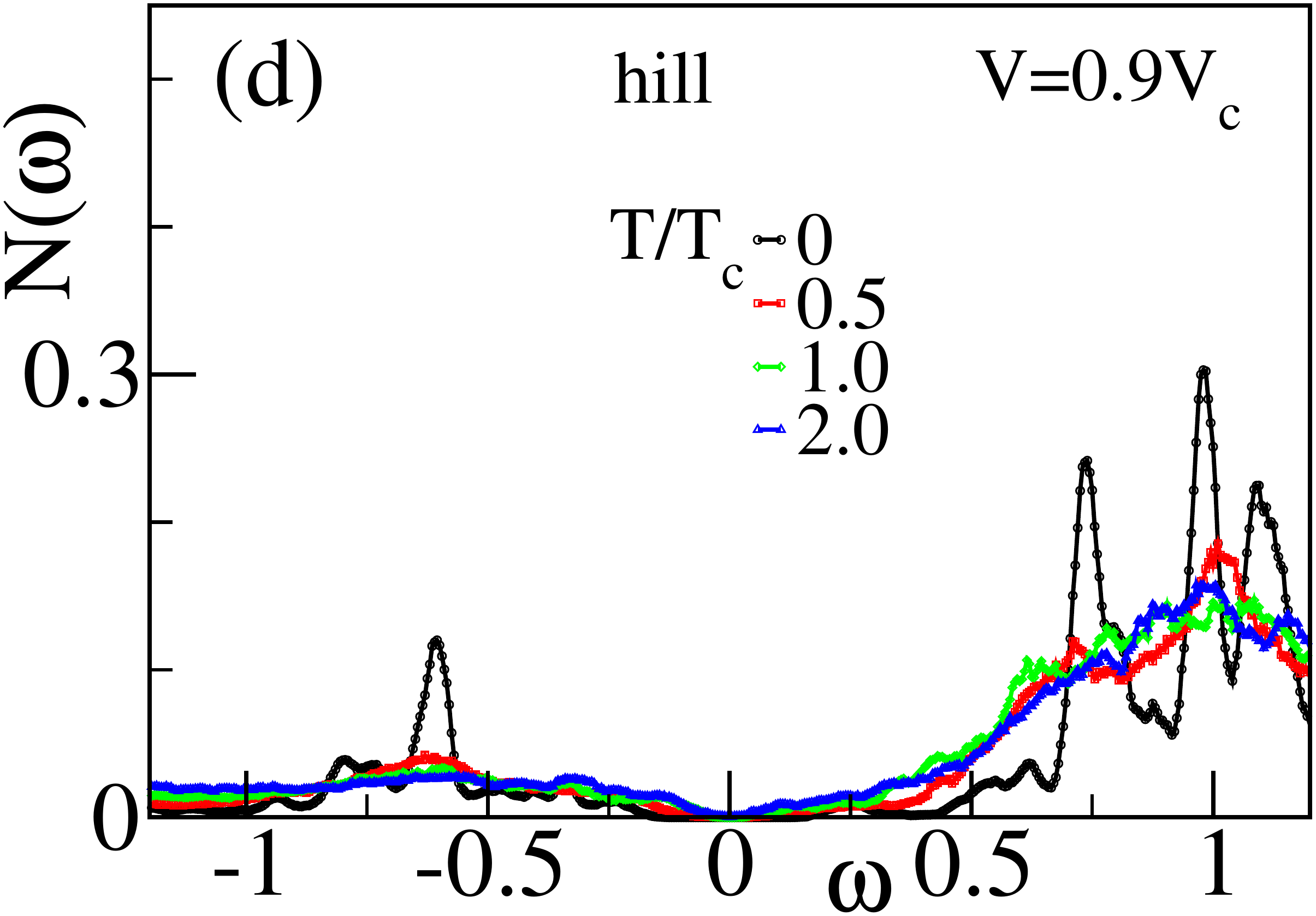}
}
\caption{Colour online:
Local DOS on a typical `plateau' site (top)
and `hill' site (bottom). The left panels (a) \& (c) are at 
$V=0.2V_{c}$,  the right
panels (b) \& (d) are for $V=0.9V_{c}$.
In each panel the low energy DOS is shown for four temperatures
$T \approx 0,~0.5T_{c},~T_{c},~2.0T_{c}$. The curve at $0.5T_{c}$
has been omitted in (b) for clarity.
}
\end{figure}

We now discuss the 
multiple scales that emerge in the strongly disordered 
superconductor, and the relevance of our results
to recent STS measurements.

(i)~{\it Multiple scales:}~In the clean limit,
weak coupling SC is characterised by only one scale, 
$T_c^0(U)$, while strong coupling brings into play
two additional \cite{scales} scales $T_g^0(U) > T_c^0(U)$ 
and $T_{pg}^0(U) > T_g^0(U)$.
This paper focuses on the weak coupling end 
where there is no gap/PG above $T_c$ at $V=0$
but disorder {\it generates} such scales.
These scales emerge due to the fragmentation of the SC
ground state with increasing disorder (the nature of the
patterns is discussed in the Supplement, and also 
\cite{ghosal-prb}). 
The inhomogeneous state leads to a spatially varying phase
and amplitude stiffness, whose distribution and 
spatial correlation dictates the thermal response.

The Supplement describes how a bond resolved phase
stiffness, $J_{ij}$, can be extracted from the 
non local pairing susceptibility
in the disordered ground state.
Focusing on nearest neighbour bonds,
at small $V$ the $J_{ij}$ are `large',
homogeneous, and $\sim {\cal O}(J_0)$, the clean 
value. At large $V$ they are strongly inhomogeneous,
with a smaller mean value $\langle J \rangle \ll J_0$.
While $\langle J \rangle$ 
(on the percolative backbone)
decides $T_c(V)$, the presence
of bonds with $J \sim J_0 \gg \langle J \rangle$,
 near the center of the SC clusters, leads to survival of 
local SC correlations to $T_{clust} \gg T_c$ as 
$V \rightarrow V_c$. The Supplement 
shows plots of the distribution $P(J_{ij},V)$ 
and a spatial map of $J_{ij}$ at stong
disorder.

Although phase fluctuations destroy global order,
the $\vert \Delta_i \vert$
survive to $T \gg T_c$. At $U=2t$ 
this sustains a gap to $T_g > T_c$ and then a PG to
a scale $T_{pg} > T_g$, ultimately closing due to
amplitude fluctuations.
The Supplement shows these scales
at $U=2t$, and also $U=4t$ to allow
extrapolation over a wider interaction window.
At $U=2t$, while $T_c \rightarrow 0$ as 
$V \rightarrow V_c$, $T_{clust}$ remains 
roughly
constant at $\sim 0.5 T_c^0$, as does $T_g$. 
$T_{pg}$ is much larger than these scales. 
At $U=4t$ $T_g \gg T_{clust}$, 
so their coincidence at $U=2t$ is accidental.
Extrapolating downward we expect that when $U \ll t$,
$T_{clust}$ will continue to be a finite fraction of
$T_c^0$, with $T_{pg} \sim T_{clust}$.

{\it Comparison with experiments:}
Our main results, {\it i.e}, 
the emergence of a $T_{clust}$, $T_g$, {\it etc}, in addition to
$T_c$, the shrinkage and fragmentation of the SC pattern with
increasing temperature, and 
the distinct thermal evolution of the STS spectra in the
SC and insulating regions,  are all in 
agreement with recent experiments.
However, there are also important differences,
arising from (i)~our parameter choice, (ii)~our approximation, 
and (iii)~the neglect of Coulomb interactions. 
(i)~Experimental spectra indicates a pseudogap
\cite{pratap-ph-diag},
rather than a hard gap above $T_c$ for $V \rightarrow V_c$.
Our exploration of the $U$ dependence suggests that
at weaker coupling such a result
would emerge from our method as well. 
Another effect of the relatively `large' coupling
that we use is 
the larger variation of local gaps between the 
SC and
insulating regions, experimentally these gaps
are comparable \cite{sacepe-loc}. 
(ii)~The neglect of quantum fluctuations
in our treatment of the A2DHM prevents access
to the correct asymptotic low temperature behaviour
for $V \rightarrow V_c$.
However, apart from the immediate vicinity
of $V_c$ the thermal fluctuations seem
to capture most of the qualitative experimental features.
(iii)~The recent observation \cite{pratap-th} 
of enhanced zero-bias conductance in 
the insulating regions is probably caused by 
additional interactions that are absent 
in our model. Also, the broad 
V-shaped background observed in the
STS spectra possibly arises from 
Coulomb interactions, and is absent in our results.

{\it Conclusions:} We have  studied the spatial signatures
of the thermal transition in a disordered s-wave superconductor
as probed by tunneling spectroscopy.
Our detailed spatial maps of the coherence and subgap features 
in the local DOS allow us to identify the distinct evolution
of the superconducting and `insulating' regions with temperature.
We point out new thermal scales, $T_{clust}$, $T_{pg}$ and $T_g$
that come into existence at strong disorder, identify their
physical origin, and quantify their dependence on disorder
and interaction strength. Recent experiments have already
indicated the existence of such scales in 2D films, our 
results provide the broader framework within which these 
results can be analysed.

{\it Acknowledgments:}
We acknowledge use of the High Performance 
Computing Cluster 
at HRI.  PM acknowledges support from a DAE-SRC 
Outstanding Research Investigator
Award, and the DST India (Athena).
We thank Amit Ghosal and P. Raychaudhuri 
for discussions.


\begin{thebibliography}{99}
\bibitem{vc-footnote} There is no SIT in the ground state
strictly within HFBdG theory. However at any finite $T$ 
phase fluctuations lead to a finite $V_c(T)$. Our $V_c$ is
an extrapolation of this quantity \cite{sit-transp}.
\bibitem{scales} Note that unlike $T_c$, the gap and pseudogap vanishing
scales are more ambiguous, particularly the later. Nevertheless, they
are useful constructs, and at least in the present case can be 
reasonably estimated.
\end{thebibliography}
\end{document}


\begin{center}
\large{\bf Supplementary material for\\
 ``Tunneling Spectroscopy Across the 
Superconductor-Insulator Thermal Transition''}
\end{center}

\begin{center}
\large{Sabyasachi Tarat and Pinaki Majumdar} 
\end{center}

\section{Computation of measurables}

In our scheme, the local density of states (LDOS)
is given by
$$
 N_{ii}(\omega) =  
\frac{1}{N} \sum_{n} 
\langle  
(|u_{n}^{i} |^{2} \delta (\omega - E_{n}) + 
|v_{n}^{i}|^{2} \delta (\omega + E_{n}) ) 
\rangle
$$
where the $u_n,v_n$ are obtained from the BdG eigenfunctions, 
$E_n$ is the positive BdG eigenvalue, and the entire expression
is thermally averaged (angular brackets) over equilibrium configurations
of the auxiliary field.
We typically average over $\sim  100$ configurations at a given
temperature.
The third
row in Fig.2 corresponds to LDOS averaged over
the interval $\delta \omega_{gap}=[0,0.15]t$, and
the fourth row to
$\delta \omega_{coh} = [0.15,0.35]t$.

The global DOS of the quasiparticles is obtained as
$N(\omega) = \langle \sum_n \delta(\omega - E_n) \rangle $, 
where now the sum runs over both positive and negative $E_n$.
The quasiparticle (QP)
gap is the minimum of $2E_n $ over
all thermal configurations at a given $T$.

\begin{figure}[b]
\centerline{
\includegraphics[width=16cm,height=4.0cm,angle=0]{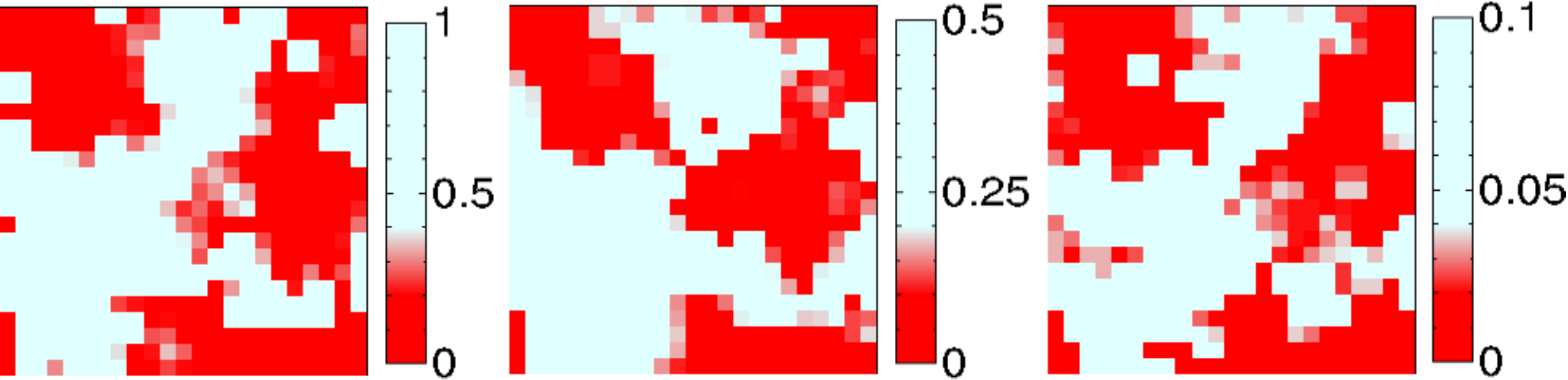}
}
\caption{Colour online: Gap maps for 3 cases, $T=0$. Left: Original calculation,
showing patches with low gaps, idential to the correlated patches.
Middle: With $V_{bare}$, scaled to lie between ($-V-2$,$V$), showing
low gap patches with same basic structure; Right: with $V_{eff}$,
increasing the contrast of the original $V$, gives similar map, with
gaps of the insulating regions raised substantially.
}
\end{figure}

\section{Cluster pattern at $T=0$}

Fig.S1 shows `gap-maps' for a particular disorder realisation
at $V=0.9V_{c}$ for three different cases: (a)~actual ground state,
(b)~ground state obtained with $V_{i}^{eff} = V_{i} - \phi_{i}$ 
and (c)~with $V_{i}$, but the potential 
scaled to lie between ($V - (U/2)n_{min}$,
$-V - (U/2)n_{max}$), with $n_{min} =0$ and $n_{max}=2$. 
The `gap'  at a site `i' is defined as the difference between the 
smallest energy $\omega_{+}>0$ at which the LDOS 
$N_{ii}(\omega) \gtrsim N_{cut}$ and the corresponding $\omega_{-}<0$,
where $N_{cut}$ is a suitably defined cutoff.

All three show the same pattern of low gap areas (coloured red) even 
though the overall scales change. Thus, superconducting clusters
form in areas that are already defined by a small local gap 
at the Fermi level. The Hartree field magnifies this
effect, further increasing the local gaps in the insulating regions,
while the $\Delta$ open up a smaller gap
on sites within the superconducting clusters.

Since regions with low local gap {\it around the chemical potential}
decide the cluster pattern, this pattern depends sensitively on 
the overall electron density in the system.

To understand the density distribution in the
system we define `hill' sites as those with $n_i \lesssim 0.4$,
valley sites as those with $n_i \gtrsim 1.6$, 
and `plateau' sites where $n_{i}$ is within
$10\%$ of $n_{avg} \sim 0.9$
The rest are `moderate' sites.
Fig.S2 shows the distribution of `plateau' sites 
inside the SC clusters (with $\Delta \gtrsim 
0.4 \Delta_{0}$) as well as outside at 
high disorder $V=0.9V_{c}$. 
The plots clearly show that a majority of the 
plateau sites lie inside the SC cluster region, 
providing a backbone for the 
formation of the clusters. 

\begin{figure}[t]
\centerline{
\includegraphics[width=16cm,height=4.0cm,angle=0]{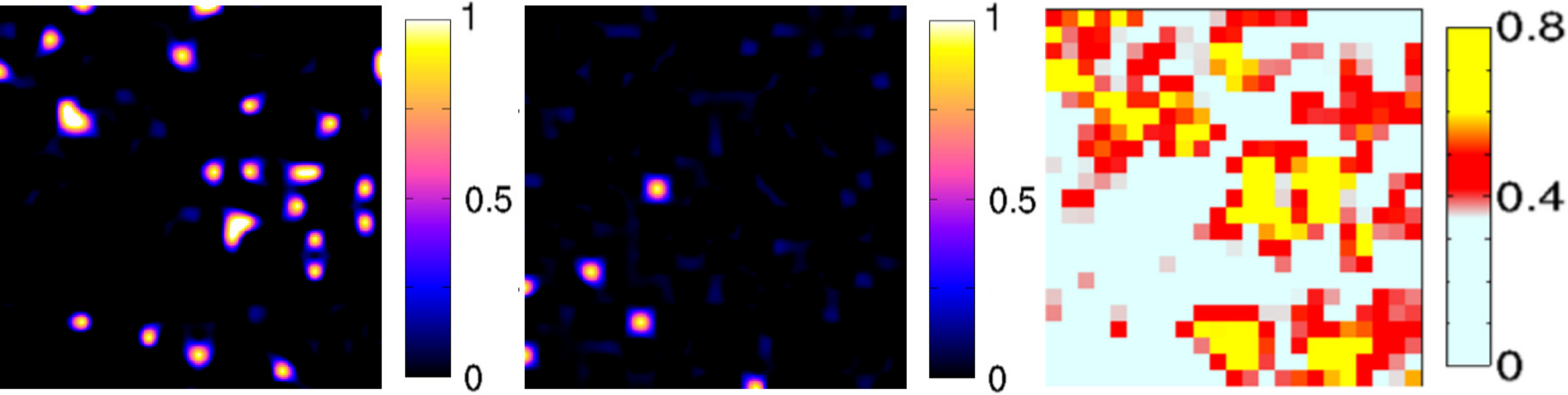}
}
\caption{Colour online: Comparison of location of `plateau' sites
(see text) inside superconducting clusters (left) with the same 
outside clusters (middle) at $V=0.9V_{c}$, $T=0$. Right figure shows 
the spatial plot of $|\Delta_{i}|$ for reference. `Pleateau' sites
form a backbone over which the superconducting clusters are formed.
In the insulating area, `hill' or `valley' sites (see text) predominate,
ruling out the formation of SC clusters.
}
\end{figure}

These only account for 
 $13\%$ of the total number of sites, but combined with the $40\%$ `moderate'
sites, they allow enough particle-hole mixing to favour the formation of 
SC clusters. The fact that {\it 
$47\%$ of sites within the SC clusters}
are of the `hill/valley' 
type shows that the SC regions are far from being `flat'. On the other 
hand, in the insulating regions, $80\%$ of the sites are of the 
`hill/valley' type, surrounding  the few isolated `plateau' site.

\section{Spatial patterns at moderate disorder}

In the main text we have discussed the strong disorder regime
where, for the system size we are studying, the cluster pattern
is quite prominent. At weaker disorder the patterns are more
ambiguous, and we discuss this $V=0.5V_c$ case here.

{\it Ground state:}
The top left panel in Fig.S3 is the 
$T=0$ pattern of
$\langle \vert \Delta_i \vert \rangle$. In contrast
to $0.9V_c$ in the main text this has 
only small regions with suppressed amplitude,
scattered in a background with moderate to
large $\langle \vert \Delta_i \vert \rangle$.
The corresponding nearest neighbour 
phase correlation 
$\Phi_i$ is almost saturated
over the system, except for small regions
which correlate with the small
$\langle \vert \Delta_i \vert \rangle$ patches.

The subgap tunneling {\it does not}
follow the trend observed at $0.9V_c$: $T_i^{gap}$ 
seems to be larger where 
$\langle \vert \Delta_i \vert \rangle$ 
is either weak or only moderately large.
There is no direct correspondence with the correlated regions.
Local gaps in the correlated region  can be
larger as well as smaller than gaps in poorly correlated 
regions, but we observe a partial recovery of the
`large $\Delta \equiv $ large gap' rule.
$T_i^{coh}$ is
large over most of the system, with larger
intensity corresponding, roughly, to larger
values of $\langle \vert \Delta_i \vert \rangle$.

\begin{figure}[t]
\centerline{
\includegraphics[width=8.1cm,height=10.8cm,angle=0]{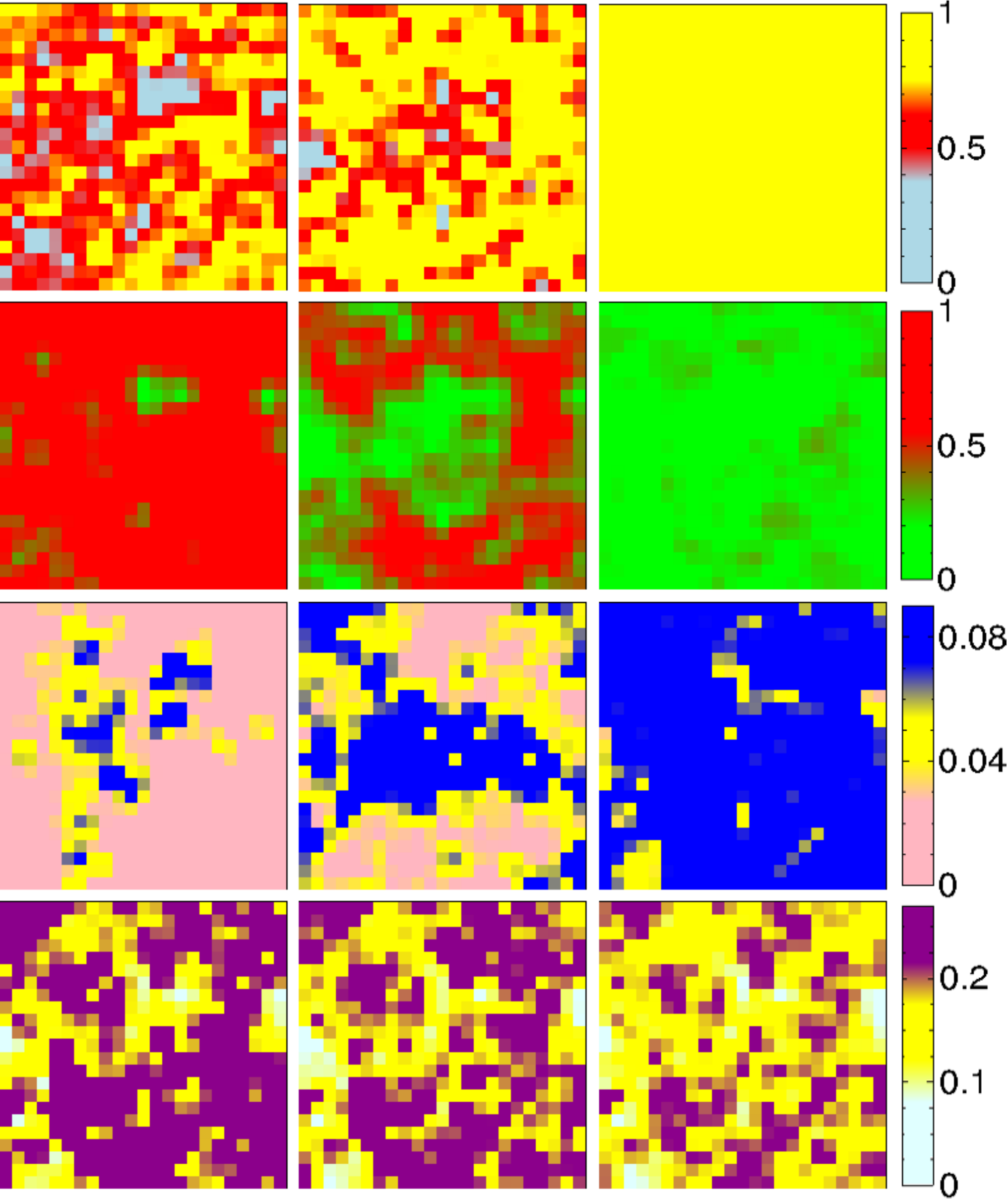}
}
\caption{Colour online:
Spatial maps at $U=2$ and $V=0.5V_c$.
First row: $\langle \vert \Delta_i \vert \rangle$,
second row: phase correlation $\Phi_i$ (see text),
third and fourth rows show the tunneling conductance
averaged over two frequency (or bias) windows $\omega_{gap}$
and $\omega_{coh}$.
Along the row: temperatures $T=0$, $0.4T_{c0}$ and $0.8T_{c0}$.
$\langle  | \Delta_i |\rangle $ at low $T$ forms
phase correlated clusters, which shrink in size as $T$ is increased.
$\langle | \Delta_i|\rangle$
is weakly inhomogeneous at low $T$, and smoothens with increasing $T$.
$\Phi_i$ decrease with $T$ and vanishes almost homogeneously
at $T_{c}$ (not shown, but between $0.5T_{c0}$ and $T_{c0}$.)
The subgap region lights up with increasing $T$, due
to the transfer of spectral weight to low frequency, while the
plot for $\omega_{coh}$ loses intensity.
}
\end{figure}

{\it Thermal evolution:}
The middle column in the right set of panels is for
$T \approx 0.5T_c$ and the right column for 
$T \approx T_c$.
 The 
$\langle \vert \Delta_i \vert \rangle$  increase on an average 
with $T$ and is uniform by $T \approx T_c$.
The loss
of phase correlation is spatially differentiated,
stronger phase correlation survives in regions where the 
$\langle \vert \Delta_i \vert \rangle $ 
is larger.
By $T_c$ the
$\langle \vert \Delta_i \vert \rangle $ 
has homogenised and phase correlations are lost
throughout, unlike the strong disorder case where correlated
patches survived to $\sim 1.3T_c$.

The subgap tunneling does not have a forceful
correspondence with the profile of 
$\langle \vert \Delta_i \vert \rangle $
or $\Phi_i$. At $T=0.5T_c$ it 
seems to show low intensity (hence
larger `gap') roughly in regions which have larger
$\langle \vert \Delta_i \vert \rangle $
 and stronger $\Phi_i$ (roughly the ring like
region excluding the center). 
At $T \sim T_c$ the intensity is large almost everywhere and
no signature of any cluster can be seen.
Large intensity in $T_i^{coh}$  at $T=0.5T_c$ similarly
has a rough correspondence with
large $\langle \vert \Delta_i \vert \rangle $
and phase correlation and only a tentative match
with $\Phi_i$ at $T \sim T_c$.
 
\section{Effective Ginzburg-Landau functional}

The interplay of `amplitude' and `phase' fluctuations in our model
is best understood in terms of a simple lattice Landau-Ginzburg model:
\begin{eqnarray}
 {\cal F}(\Delta) & =& \sum_{i \neq j} J_{ij} \Delta_{i} \Delta^{*}_{j} + 
\sum_{i} (a_i |\Delta_{i}|^{2} +  b_{i} |\Delta_{i}|^{4} )
\end{eqnarray}
The $a_i$ and $b_i$, crudely,  control the $\vert \Delta_i \vert$
while $J_{ij}$ 
determine a bond coupling between the `i' and `j' sites
(not necessarily nearest neighbours). 
These parameters are 
in general temperature dependent, and the clean problem at
$U=2t$  involves 
renormalisation of all these parameters with temperature.
We focus here on understanding the strong disorder 
regime $V \gtrsim 0.75V_{c}$
and the thermal evolution of the superconducting clusters only
over a small temperature window above the ground state (the
$T_c$ here is small). 
Such a situation allows us to
ignore the thermal renormalisation of the GL parameters as a first
approximation.

We calculate $J_{ij}$ via a perturbative expansion of the
energy in $\Delta_i$ around the disordered ground state.
This leads to:
$
J_{ij} \sim {1 \over \beta} \sum_{n} 
G_{ij}(i \omega_n) G_{ij}(- i \omega_n)
$
where  $G_{ij}(i \omega_n)$
is the electronic Greens function computed in the background
defined by $V_i$ and $\phi_i$.
We compute $G_{ij}$ exactly in the $\{ V_i + \phi_i\}$
background. 
The $a_i$ and $b_i$ can be found by fitting the local 
amplitude
distributions to the given form, but we concentrate on $J_{ij}$ here.

Fig.S4 compares the phase correlations obtained from 
the bond disordered 
XY model, above, with that from the full Monte Carlo at $V=0.8V_{c}$.
We find a reasonable match.

Fig.S5 shows the distribution of the nearest neighbour $J_{ij}$ at various
disorder values and a spatial map of the same at strong disorder, $V=0.9V_{c}$. 
The distributions broaden with increasing disorder, 
and their means shift to lower values, but 
a finite proportion of sites have $J \sim J_{0}$ even at large disorder, 
where $J_{0}$ is the clean value $\sim 0.023$. These correlations thus 
only vanish when $T \sim J \Delta^{2}$, where $\Delta$ is typically 
a large fraction of the clean $\Delta_{0}$, explaining why $T_{clust} >> T_{c}$
even at large disorder. A comparison of the $J_{ij}$ map with the low temperature correlation 
map in Fig.S4 clearly shows that SC clusters generally
have larger $J_{ij}$, with the middle of the clusters having the largest
stiffnesses. Thus, under thermal evolution, these regions are the last 
to vanish.

\begin{figure}[t]
\centerline{
\includegraphics[width=14cm,height=4.5cm,angle=0]{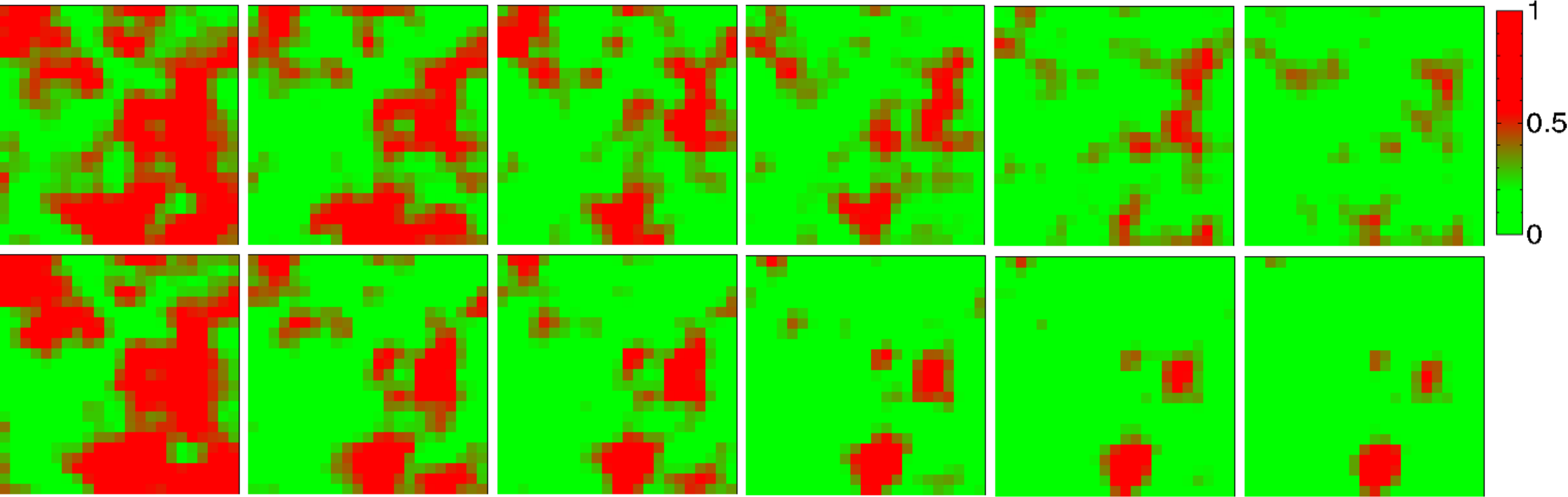}
}
\caption{Colour online: Comparison of correlated patches for $V=0.8V_{c}$ for
successive $T$ points using the full Monte Carlo (top) and the simplified XY
model (bottom). 
The basic phenomenon of an island pattern at low $T$, and shrinkage of these
clusters with increasing temperature 
is well captured by the simplified model.
}
\end{figure}
\begin{figure}[t]
\centerline{
\includegraphics[width=12cm,height=5.0cm,angle=0]{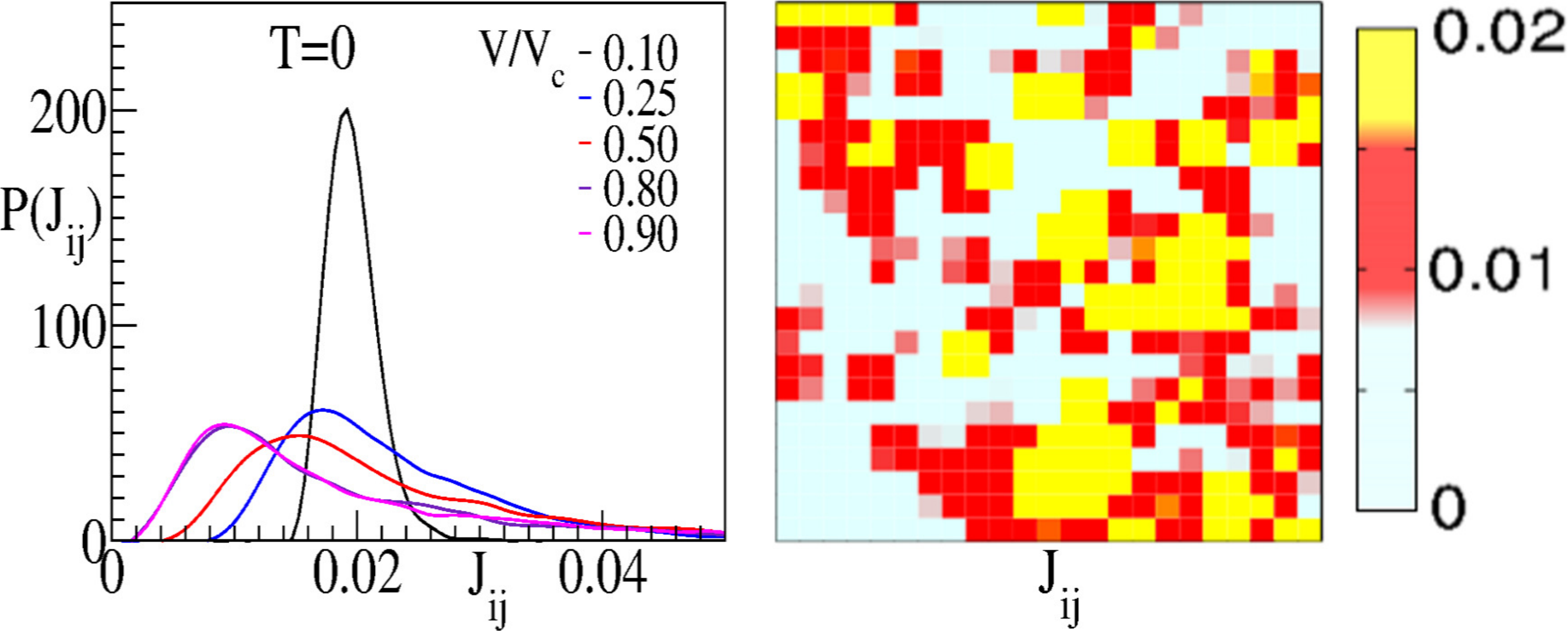}
}
\caption{Left: Distribution of nearest neighbour $J_{ij}$ at different disorder,
showing successive broadening with disorder, but even at strong disorder, 
a finite number of sites have $J \sim O(J_{0})$. Right: Spatial map of $J_{ij}$,
at $V=0.9V_{c}$, showing that the large values of $J_{ij}$ correspond to
 the centres of the SC clusters.
}
\end{figure}

\section{Dependence of temperature scales on disorder and coupling}

Fig.S6 shows the phase diagrams for $U=2$ and $4$. The tails emphasize the fact 
that in principle, the $T=0$ state is always superconducting within our scheme. 
The increased Hartree field at stronger coupling increases the effective 
disorder $V_{i} - \phi_{i}$, leading to a smaller $V_{c}$. This also leads
to a faster fragmentation with disorder, and a more differentiated $J_{ij}$
distribution, which results in a faster decline in $T_{clust}$ at $U=4$ compared 
to that at $U=2$. The gap vanishing scale, on the other hand, increases 
considerably with coupling, and at $U=4$, it is greater than the clean $T_c^0$
at all disorder values considered here. Thus, the similar magnitudes for 
$T_{clust}$
and $T_{g}$ observed in experiments is only seen at weak coupling ($U=2$), and 
is not true in general.

\begin{figure}[h]
\centerline{
\includegraphics[width=5cm,height=4.2cm,angle=0]{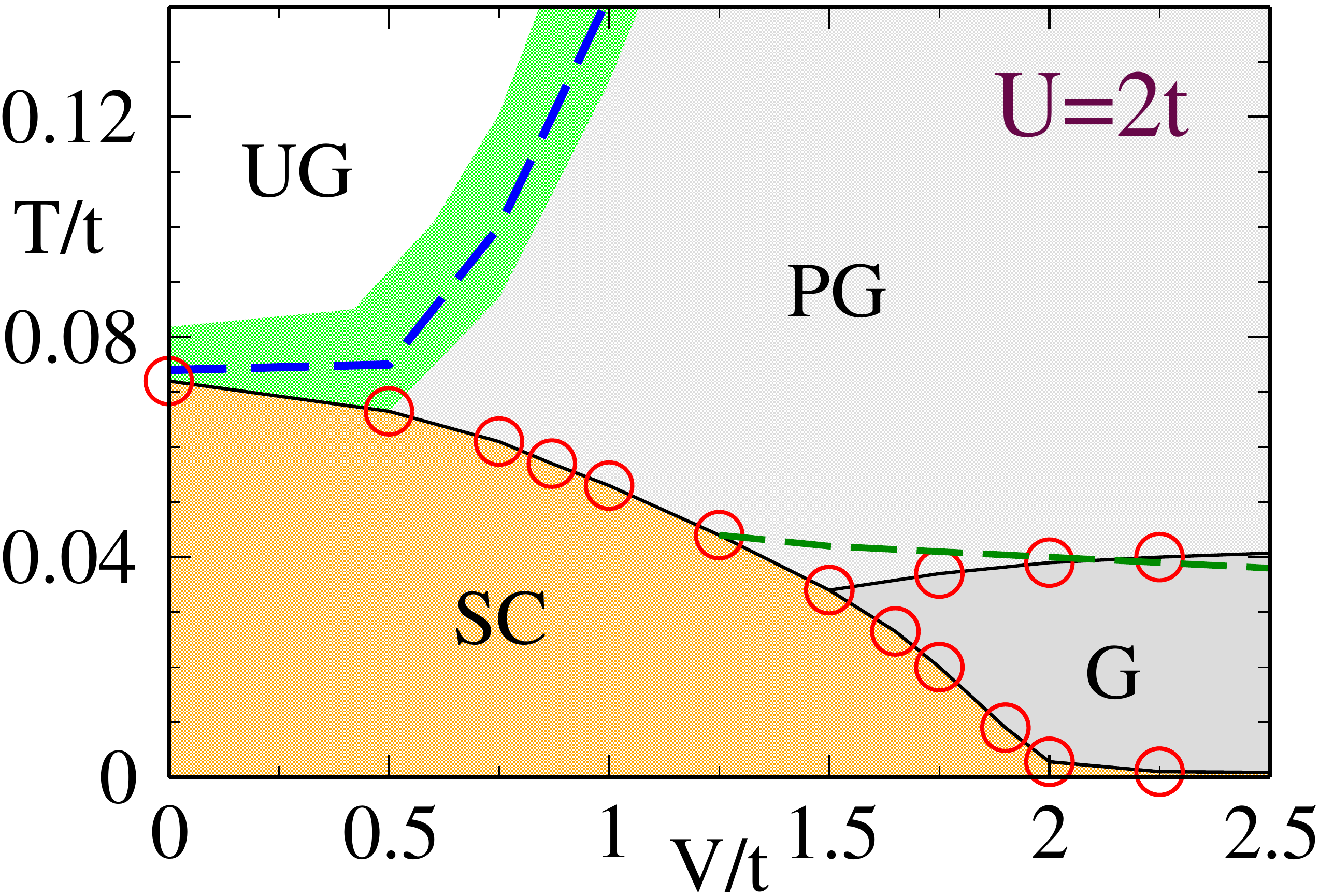}
\includegraphics[width=5cm,height=4.2cm,angle=0]{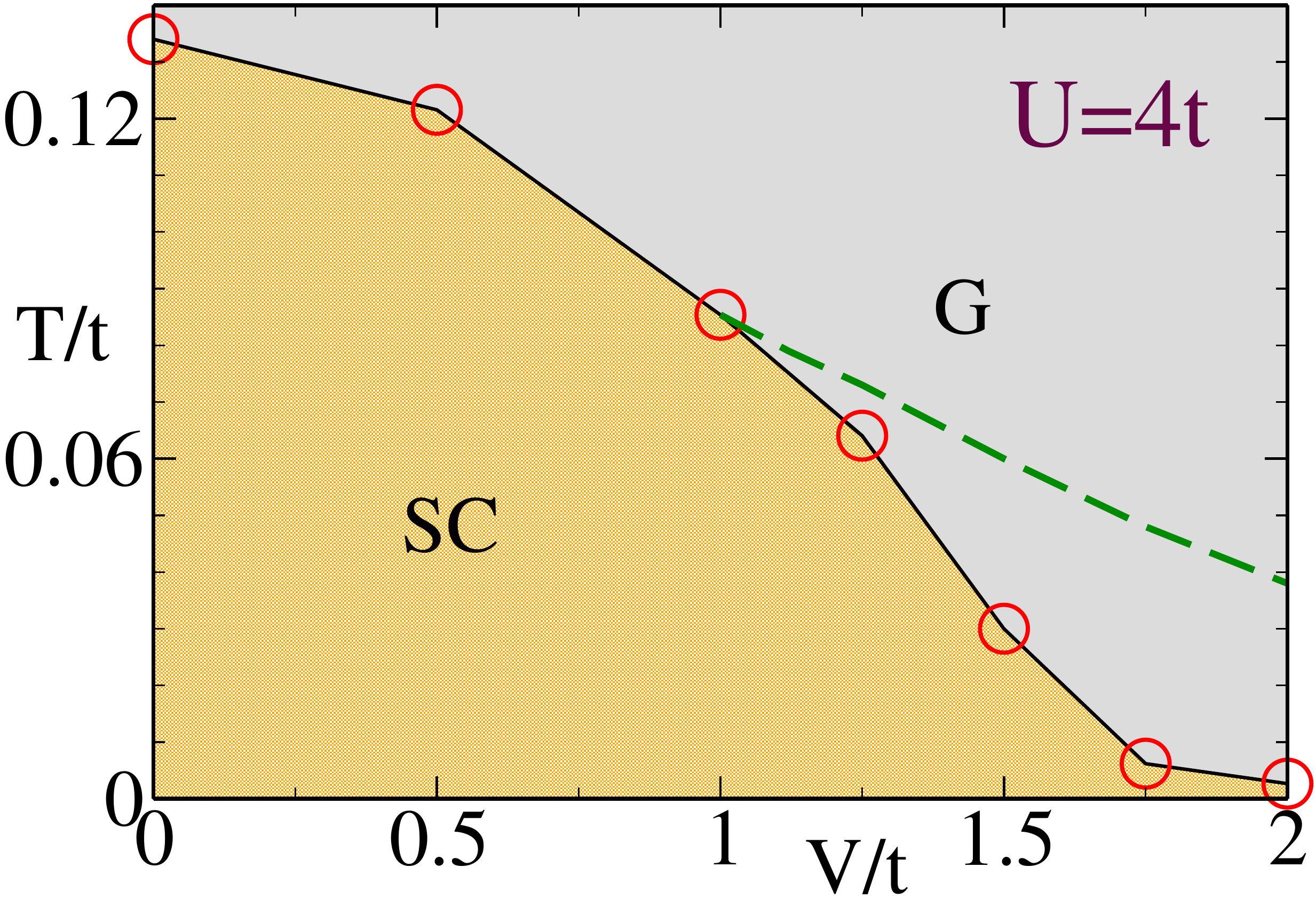}
}
\caption{Colour online: Phase diagrams showing superconducting (SC),
gapped (G), pseudogapped (PG) and ungapped (UG) phases, and the cluster
vanishing scale $T_{clust}$ (green dashed lines) at $U=2$ and $4$ (see text).
}
\end{figure}